\documentclass[twocolumn,superscriptaddress]{revtex4-2}
\usepackage{xcolor}
\usepackage{graphicx}
\usepackage{dcolumn}
\usepackage{bm}
\usepackage{amsmath}
\usepackage{amsfonts}
\usepackage{multirow}
\usepackage[colorlinks=true, a4paper=true, pdfstartview=FitV,
linkcolor=blue, citecolor=blue, urlcolor=blue]{hyperref}

\begin{document}
\definecolor{mygreen}{HTML}{006E28}
\newcommand{\ped}[1]{{\color{mygreen}\textbf{?PED:}  #1}}
\newcommand{\trom}[1]{{\color{red}\textbf{?TR:} \color{red} #1}}
\newcommand{\ya}[1]{\textbf{?YS:} {\color{blue} #1}}
\newcommand{\aw}[1]{{\color{magenta}\textbf{?AW:}  #1}}


\title{Dissecting normal modes of vibration on vortices in Ginzburg-Landau superconductors}

\author{A. Alonso-Izquierdo}
\affiliation{
 Departamento de Matemática Aplicada, University of Salamanca, Casas del Parque 2, 37008 Salamanca, Spain
}

\affiliation{
	IUFFyM, University of Salamanca, Plaza de la Merced 1, 37008 Salamanca, Spain
}

\author{D. Migu\'elez-Caballero}
\affiliation{Departamento de Física Teórica, Atómica y Óptica and Laboratory for Disruptive Interdisciplinary Science (LaDIS), Universidad de Valladolid, 47011 Valladolid, Spain
}

\begin{abstract}
The structure of the normal modes of vibration of rotationally invariant $n$-vortices in the Ginzburg-Landau/Abelian Higgs model is completely unveiled for any value of the coupling constant.  
\end{abstract}

\maketitle

\section{\label{sec:intro} Introduction}

Filaments carrying $n$ quanta of magnetic flux ($n$-vortices) were first discovered by Abrikosov within the framework of the Ginzburg-Landau theory of Type II superconductors \cite{Abrikosov}. These magnetic flux tubes reappeared in the relativistic setting of the Abelian Higgs model in the seminal work of Nielsen and Olesen \cite{Nielsen}. Such solutions are also of cosmological significance, where they manifest as cosmic strings \cite{Vilenkin}. For the critical value of the coupling constant $\lambda=1$ (the transition point between Type I and Type II superconductivity) it was demonstrated that the vortices satisfy first-order BPS (Bogomolny-Prasad-Sommerfield) PDE \cite{Bogomolny,Prasad}. This endows the vortices with special properties \cite{Taubes}, particularly, the absence of interactions among them, allowing the vortices to move freely in space. This implies the existence of $2 n$ linearly independent zero modes of fluctuation. This result was rigorously established by E. Weinberg \cite{Weinberg} through a generalization of the index theorem for elliptic operators. The existence of these zero modes is central to the analysis of low-energy vortex dynamics, which can be understood as geodesic motion in the moduli space of BPS vortex solutions \cite{Ruback,Burzlaff,Manton}. In addition to these zero modes, the second-order small fluctuation operator associated with the self-dual vortices includes massive discrete modes. Derrick-type bound states, where the vortex size oscillates periodically, were identified by Goodman and Hindmarsh \cite{Hindmarsh}. These modes exhibit the same angular dependence as the static $n$-vortex solution. Other eigenfluctuations, however, do not follow this form. Indeed, the complete set of normal modes for the rotationally invariant $n$-vortex at the critical value $\lambda=1$ was unveiled in \cite{AlGarGuil,AlGarGuil2}, where the hidden supersymmetry of the spectral problem associated with the self-dual vortex fluctuation operator was exploited. For the 1-vortex, it was shown that there are two zero modes, one Derrick-type mode with frequency $\omega^2 = 0.777476$, and a continuous spectrum starting at the threshold value $\omega^2=1$. In contrast, the rotationally invariant 2-vortex exhibits a richer structure, involving four zero modes, one Derrick-type mode with frequency $\omega^2 = 0.53859$, a doubly degenerate shape mode with $\omega^2 = 0.97303$, and a continuous spectrum beginning at $\omega^2=1$. As the vorticity increases, the spectrum becomes more intricate.

The analysis of the spectral problem for the second-order small fluctuation operator associated with a solution is often essential to understanding its properties. For instance, stability is ensured if the eigenvalues are non-negative. However, the significance of eigenfluctuations extends beyond stability. Recent studies have demonstrated that the dynamics of two self-dual vortices are drastically altered when they are excited \cite{Alonso2024a,Alonso2024b}. In the absence of excitation, the vortices do not interact. However, when they vibrate in phase (due to the aforementioned Derrick-type mode), an attractive force emerges; conversely, when they vibrate out of phase, a repulsive force arises. Additionally, a resonant energy transfer mechanism can occur during vortex collisions, facilitating energy exchange between different eigenmodes and leading to chaotic vortex dynamics \cite{Krusch}. Other works emphasize the role of these vibration modes in the context of topological defects in cosmology \cite{Arodz1,Arodz2,Kojo,Alonso2024c}. Notably, interesting effects of the massive states on vortex cores have also been identified in the superfluid phase of the isotope ${}^3{\rm He}$ \cite{Kopnin}. Moreover, the bound modes play a key role in estimating quantum corrections to the mass of topological defects \cite{Rajaraman}.

Despite the importance of the vibrational modes, the full spectral structure of the rotationally invariant $n$-vortex beyond the critical point $\lambda = 1$ has remained largely unexplored. For $\lambda \neq 1$, the supersymmetry structure present at the critical value is lost, necessitating a different mathematical approach to attack the problem. In this paper, we present a comprehensive description of the spectral structure of the vortex fluctuation operator as a function of the coupling constant $\lambda$. One application of this study is in understanding vortex dynamics in the Abelian-Higss model when $\lambda\neq 1$. It is well known that in Type I superconductors ($\lambda < 1$), vortices attract each other, whereas in Type II superconductors ($\lambda > 1$), vortices repel. However, as discussed earlier, this interaction scheme changes drastically when the vortices are in an excited state. Excited vortices in Type I materials could exhibit repulsive behavior, while vibrations could stabilize $n$-vortices in Type II superconductors.

\section{\label{sec:dynamics} The Abelian Higss model: Vortex fluctuations}

The action of the Abelian Higgs model describes the minimal coupling between a $U(1)$-gauge field and a charged scalar field in a phase where the gauge symmetry is spontaneously broken. In terms of non-dimensional coordinates, couplings and fields, the action functional for this relativistic system in $\mathbb{R}^{1,2}$ Minkowski space-time reads:
\[
S=\int d^3 x \left[ -\frac{1}{4} F_{\mu\nu}F^{\mu \nu} + \frac{1}{2} \overline{D_\mu \phi} D^\mu \phi -\frac{\lambda}{8} (\,\overline{\phi}\, \phi-1)^2 \right] \, .
\label{action1}
\]
The main ingredients are one complex scalar field, $\phi(x)=\phi_1(x)+i\phi_2(x)$, the vector potential $A_\mu(x)=(A_0(x),A_1(x),A_2(x))$, the covariant derivative $D_\mu \phi(x) = (\partial_\mu -i A_\mu(x))\phi(x)$ and the electromagnetic field tensor $F_{\mu\nu}(x)=\partial_\mu A_\nu(x) - \partial_\nu A_\mu(x)$. We choose the metric tensor in Minkowski space in the
form $g_{\mu\nu}={\rm diag}(1,-1,-1),$ with $\mu,\nu=0,1,2$, and use the usual Einstein repeated index convention. In the temporal gauge $A_0=0$, the second order PDE for the complex field $\phi$ and the spatial components of the vector field $A_\mu$ are given by 	
\begin{eqnarray}
&& \frac{1}{2}\partial_0^2 \phi - \frac{1}{2}D_jD_j\phi = - \frac{\lambda}{4} \phi (\,\overline{\phi}\,\phi -1) \hspace{0.9cm}, \label{genedo1} \\
&& \partial_0^2 A_j  - \partial_k F_{kj} = - \frac{i}{2} \Big[ \overline{\phi} \, D_j \phi -  \overline{D_j \phi} \,\phi \Big] \hspace{0.5cm}. \label{genedo2}
\end{eqnarray}
The equations (\ref{genedo1}) and (\ref{genedo2}) must be supplemented by the Gauss law $\partial_{0i}A_i = - \frac{i}{2} ( \overline{\phi} \, \partial_0 \phi -\phi  \,\overline{\partial_0 \phi} )$, which is automatically satisfied for static solutions. Rotationally invariant $n$-vortex solutions with quantized magnetic flux $\Phi=\frac{1}{2\pi} \int_{\mathbb{R}^2} d^2 x F_{12} =n$ can be identified by imposing the radial gauge condition $A_r=0$ and the ansatz 
\begin{equation}
\phi(r,\theta) = f_n(r) e^{in\theta} \hspace{0.3cm},\hspace{0.3cm} r A_\theta(r,\theta) = n \beta_n(r) \hspace{0.2cm} . \label{ansatz}
\end{equation}
where we have used spatial polar coordinates $(r,\theta)$. 

The radial profile functions $f_n(r)$ and $\beta_n(r)$ must comply with the differential equations
\begin{equation}\label{edof}
\begin{aligned} & \frac{d^2 f_n}{dr^2} + \frac{1}{r} \frac{df_n}{dr} - \frac{n^2 (1-\beta_n)^2 f_n}{r^2} + \frac{\lambda}{2} f_n (1-f_n^2) = 0,  \\
& \frac{d^2 \beta_n}{dr^2} - \frac{1}{r} \frac{d\beta_n}{dr} + (1-\beta_n) f_n^2 = 0 \, \, . 
\end{aligned}
\end{equation}
Near $r=0$, these functions behave as $f_n(r) \sim d_0 r^n$ and $\beta_n(r) = c_0 r^2$ for some constants $d_0,c_0\in \mathbb{R}$, while their asymptotic behavior is given by the relations $\lim_{r\rightarrow \infty} f_n(r)= 1$ and $\lim_{r\rightarrow \infty} \beta_n(r) = 1$. For intermediate values of $r$ the equations (\ref{edof}) must be solved numerically. The scalar and vector fields corresponding to this solution will be respectively represented as $\psi(\vec{x})=\psi_1(\vec{x}) + i \, \psi_2(\vec{x})$ and $V(\vec{x})=(V_1(\vec{x}),V_2(\vec{x}))$ with $\vec{x}=(x_1,x_2)$ in the subsequent formulas. 

The fluctuations of the vortex solution will be denoted as $\varphi(\vec{x})=\varphi_1(\vec{x})+i\varphi_2(\vec{x})$ and $a(\vec{x})= (a_1(\vec{x}),a_2(\vec{x}))$. To discard pure gauge fluctuations, we impose the \textit{background gauge} condition \cite{Hindmarsh,AlGarGuil,AlGarGuil2}
\begin{equation}
\partial_k a_k( \vec{x})-(\,\psi_1( \vec{x})\, \varphi_2( \vec{x})-\psi_2( \vec{x})\,\varphi_1( \vec{x})\,)=0
\label{backgroundgauge}
\end{equation}
as the gauge fixing condition on the fluctuation modes. If we assemble the perturbation fields as
\[
\xi(\vec{x})=\left( \begin{array}{c c c c}a_1(\vec{x}) & a_2(\vec{x}) & \varphi_1(\vec{x}) & \varphi_2(\vec{x}) \end{array} \right)^t
\]
the normal modes of vibration of a $n$-vortex solution are determined by the spectral condition 
\begin{equation}
{\cal H}^+ \xi_\mu(\vec{x}) =\omega_\mu^2 \, \xi_\mu(\vec{x}) \,\, , \label{spectralproblem}
\end{equation}
where $\mu$ is a label used to enumerate the eigenfunctions and eigenvalues. ${\cal H}^+$ is the second-order vortex small fluctuation operator
\begin{widetext}
	\begin{equation}
	{\cal H}^+= \left( \begin{array}{cccc}
	-\Delta + |\psi|^2 & 0 & -2D_1 \psi_2 & 2 D_1 \psi_1 \\
	0 & -\Delta +|\psi|^2 & -2 D_2 \psi_2 & 2 D_2 \psi_1 \\
	-2 D_1 \psi_2 & -2 D_2\psi_2 & -\Delta + \frac{3\lambda}{2} \psi_1^2 + (1+\frac{\lambda}{2})\psi_2^2 -\frac{\lambda}{2}  +V_kV_k & -2 V_k \partial_k + (\lambda-1)\psi_1\psi_2\\
	2D_1\psi_1 & 2 D_2 \psi_1 & 2V_k \partial_k + (\lambda-1)\psi_1\psi_2 & -\Delta + (1+\frac{\lambda}{2})\psi_1^2 +\frac{3\lambda}{2} \psi_2^2  -\frac{\lambda}{2} + V_kV_k
	\end{array} \right) \quad \label{hessianoperator}
	\end{equation}
\end{widetext}
where $D_i\psi_j=\partial_i\psi_j+\epsilon^{jk}V_i\psi_k$. The operator (\ref{hessianoperator}) comes from linearizing the field equations (in the background gauge) around the vortices. The fluctuation vectors $\xi(\vec{x})$ belong in general to a rigged Hilbert space, such that there exist square integrable eigenfunctions $\xi_j(\vec{x})\in L^2(\mathbb{R}^2)\oplus \mathbb{R}^4$ belonging to the discrete spectrum, for which the norm $\|\xi(\vec{x})\|= \int_{\mathbb{R}^2} d^2x [ \|a(\vec{x})\|^2  + |\varphi(\vec{x})|^2 ] < +\infty$ is bounded, together with continuous spectrum eigenfunctions $\xi_\nu(\vec{x})$ with $\nu$ ranging in a dense set. In this section we have used the convention $A_1=A_r \cos \theta - A_\theta \sin \theta$ and $A_2=A_r \sin \theta + A_\theta \cos \theta$.

\section{Normal modes of vibration of vortices}

In this section, we will investigate the normal modes of vibration, $\xi(\vec{x},n,k)$, of a rotationally invariant $n$-vortex. To achieve this, we will distinguish between two types of fluctuations, each leading to a distinct spectral problem. The first class of eigenfunctions follows the form
\begin{equation}
\xi(\vec{x},n,0)= \left( \begin{array}{c} v^{(n,0)}(r) \, \sin \theta  \\ - v^{(n,0)}(r) \,\cos \theta  \\ u^{(n,0)}(r) \, \cos(n\theta) \\ u^{(n,0)}(r) \, \sin(n\theta) 
\end{array} \right) \label{eigenfunctionk0}
\end{equation}
and describes Derrick-type modes, which preserves the same angular dependence as the $n$-vortex solution (\ref{ansatz}). We associate these modes with an integer index $k=0$. The functions $v^{(n,0)}(r)$ and $u^{(n,0)}(r)$ in (\ref{eigenfunctionk0}), giving respectively the radial profile of the vector and scalar fluctuations, are determined by the spectral problem
\begin{widetext}
\begin{eqnarray}
	&& - \frac{d^2v^{(n,0)}}{dr^2} - \frac{1}{r} \frac{dv^{(n,0)}}{dr} + \Big( \frac{1}{r^2} + f_n^2(r) \Big) v^{(n,0)}(r) + \frac{2n}{r} (1-\beta_n(r)) f_n(r) u^{(n,0)}(r) = \omega_n^2 \, v^{(n,0)}(r) \hspace{0.2cm} , \label{spectralproblem0} \\
	&& - \frac{d^2u^{(n,0)}}{dr^2} - \frac{1}{r} \frac{du^{(n,0)}}{dr} + \Big( \frac{n^2 (1-\beta_n(r))^2}{r^2} + \frac{3 \lambda}{2} f_n^2(r) - \frac{\lambda}{2} \Big) u^{(n,0)}(r) + \frac{2n}{r} (1-\beta_n(r)) f_n(r) v^{(n,0)}(r) = \omega_n^2 u^{(n,0)}(r).  \nonumber
\end{eqnarray}
The second class of eigenfunctions is characterized by the expression
\begin{equation}
\xi_\mu(\vec{x},n,k) =  \left( \begin{array}{c} \cos(k \theta) \sin(\theta) \left[ \frac{dv^{(n,k)}}{dr} -r f_n(r) w^{(n,k)}(r) \right] - \frac{k}{r}   \sin (k\theta) \cos \theta \, v^{(n,k)}(r)\\ - \cos(k \theta) \cos(\theta) \left[ \frac{dv^{(n,k)}}{dr} -r f_n(r) w^{(n,k)}(r) \right] - \frac{k}{r}  \sin (k\theta) \sin \theta \, v^{(n,k)}(r) \\ \cos(k\theta)\cos(n\theta) \, u^{(n,k)}(r) + k \, \sin(k\theta) \sin(n\theta) \, w^{(n,k)}(r) \\ \cos(k\theta)\sin(n\theta) \, u^{(n,k)}(r) - k \, \sin(k\theta) \cos(n\theta) \, w^{(n,k)}(r) 
\end{array} \right) \hspace{0.5cm}, \label{eigenfunctionk}
\end{equation}
where $k=1,2,\dots$. Now, the spectral problem 
\begin{eqnarray}
	&& - \frac{d^2 v^{(n,k)}}{dr^2} - \frac{1}{r} \frac{dv^{(n,k)}}{dr}+ \Big[ \frac{k^2}{r^2} + f_n^2(r) \Big] v^{(n,k)}(r) + 2(f_n(r) +r f_n'(r)) w^{(n,k)}(r) = \omega_n^2 v^{(n,k)}(r) \hspace{0.2cm}, \nonumber \\
	&& - \frac{d^2 u^{(n,k)}}{dr^2} - \frac{1}{r} \frac{du^{(n,k)}}{dr}+ \Big[ \frac{k^2}{r^2} + \frac{n^2(1-\beta_n(r))^2}{r^2} + \frac{3\lambda}{2} f_n^2(r) - \frac{\lambda}{2} \Big] u^{(n,k)}(r)- \frac{2n(1-\beta_n(r))(k^2 + r^2 f_n^2(r))}{r^2} w^{(n,k)}(r)  + \nonumber  \\
	&& \hspace{0.5cm}+ \frac{2n(1-\beta_n(r))f_n(r)}{r} \frac{dv^{(n,k)}}{dr} = \omega_n^2 u^{(n,k)}(r) \hspace{0.2cm}, \label{spectralproblem1} \\
	&& - \frac{d^2 w^{(n,k)}}{dr^2} - \frac{1}{r} \frac{dw^{(n,k)}}{dr}+ \Big[ \frac{k^2}{r^2} + \frac{n^2(1-\beta_n(r))^2}{r^2} + f_n^2(r) + \frac{\lambda}{2} f_n^2(r) - \frac{\lambda}{2} \Big] w^{(n,k)}(r)- \frac{2n(1-\beta_n(r))}{r^2} u^{(n,k)}(r) + \nonumber \\
	&& \hspace{0.5cm}+ \frac{2f_n'(r)}{r} v^{(n,k)}(r) = \omega_n^2 w^{(n,k)}(r) \hspace{0.2cm}, \nonumber
\end{eqnarray}
\end{widetext}
fixes the radial profiles $v^{(n,k)}(r)$, $u^{(n,k)}(r)$ and $w^{(n,k)}(r)$ of the fluctuations for distinct vorticities $n$ and angular momenta $k$. The eigenfunctions (\ref{eigenfunctionk0}) and (\ref{eigenfunctionk}) automatically verify the gauge condition (\ref{backgroundgauge}). Notably, when either (\ref{eigenfunctionk0}) or (\ref{eigenfunctionk}) is substituted into the spectral problem (\ref{spectralproblem}), some of the resulting equations become combinations of others, ensuring the consistency of the procedure. For $k\geq 1$, a second linearly independent eigenfunction $\chi_\mu(\vec{x},n,k)$ can be constructed simply by making the changes $\cos(k\theta)\rightarrow \sin(k\theta)$ and $\sin(k\theta)\rightarrow -\cos(k\theta)$ in (\ref{eigenfunctionk}), which satisfies the same spectral problem (\ref{spectralproblem1}). Therefore, eigenvalues with $k\geq 1$ are doubly degenerate.

To gain insight into the behavior of the eigenfunctions near $r=0$, we plug the expansion $v^{(n,k)}(r)= r^s \sum_{i=0}^\infty v_i r^i$, $u^{(n,k)}(r)= r^t \sum_{i=0}^\infty u_i r^i$ and $w^{(n,k)}(r)= r^m \sum_{i=0}^\infty w_i r^i$ into the radial spectral problems (\ref{spectralproblem0}) and (\ref{spectralproblem1}). The following results are obtained:

\vspace{0.1cm}

\noindent (1) For $k=0$, consistency conditions at the lowest orders require that $s=1$, $t=n$ with $v_0,u_0\in \mathbb{R}$ and $v_1,u_1=0$.

\vspace{0.1cm}

\noindent (2) For $k>0$, consistency can be achieved by imposing one of two sets of conditions: 

\vspace{0.1cm}

(2.1) We can choose $s=k$ and $t=m=n-k$, with $v_0,w_0\in\mathbb{R}$, subject to the additional constraint $u_0=(n-k)w_0$ and $v_1,u_2,w_1=0$. This type of eigenfunctions corresponds to a finite number of modes because, in order to avoid singularities at the vortex center, we must impose $k=1,2,\dots, n$. At the critical point $\lambda=1$, these eigenfunctions represent the zero modes, see \cite{AlGarGuil,AlGarGuil2}. 

(2.2)Alternatively, we can set $s=k$ and $t=m=n+k$ with the additional condition $u_0=-kw_0+d_0 v_0$ and $v_0,w_0\in \mathbb{R}$, while $u_1=v_1=w_1=0$. These eigenfunctions correspond to the shape modes in the self-dual case, see \cite{AlGarGuil,AlGarGuil2}.

\begin{figure}
	\centering
	\includegraphics[height=5cm]{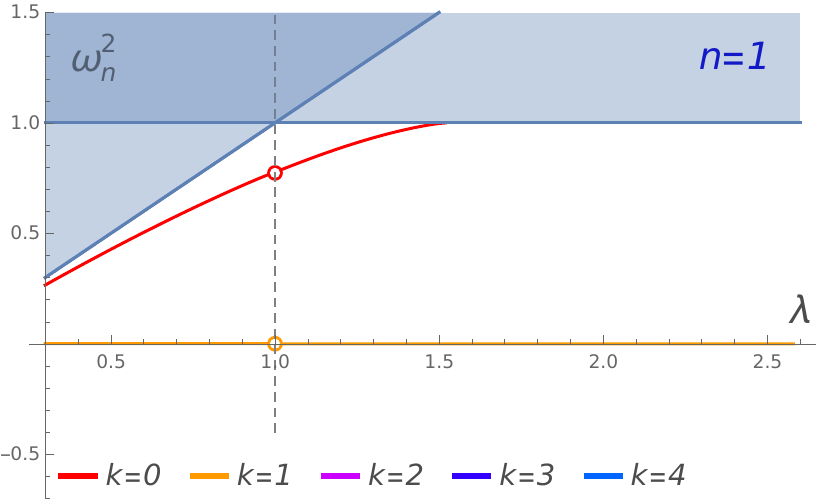}
	\caption{Spectral structure for the $1$-vortex as a function of the coupling constant $\lambda$. }  
	\label{fig:espectra1}
\end{figure}

\begin{figure}
	\centering
	\includegraphics[height=5cm]{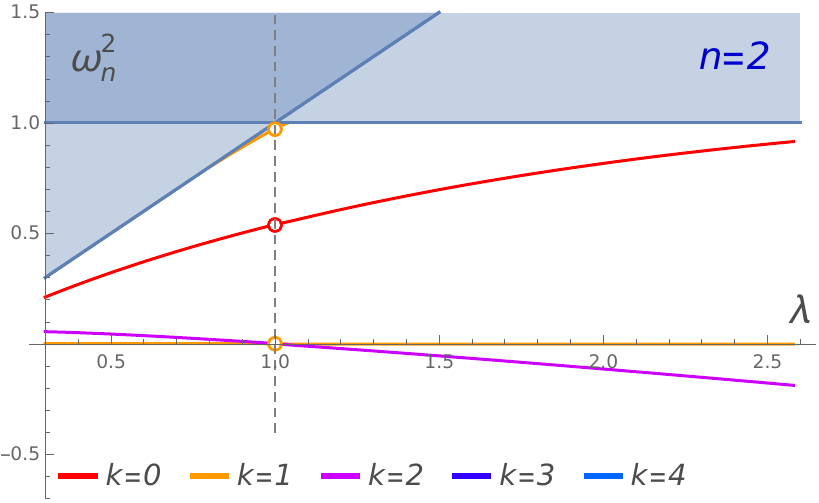} 
	\caption{Spectral structure for the rotationally invariant $2$-vortex as a function of the coupling constant $\lambda$. }  
	\label{fig:espectra2}
\end{figure}

\begin{figure}
	\centering
	\includegraphics[height=5cm]{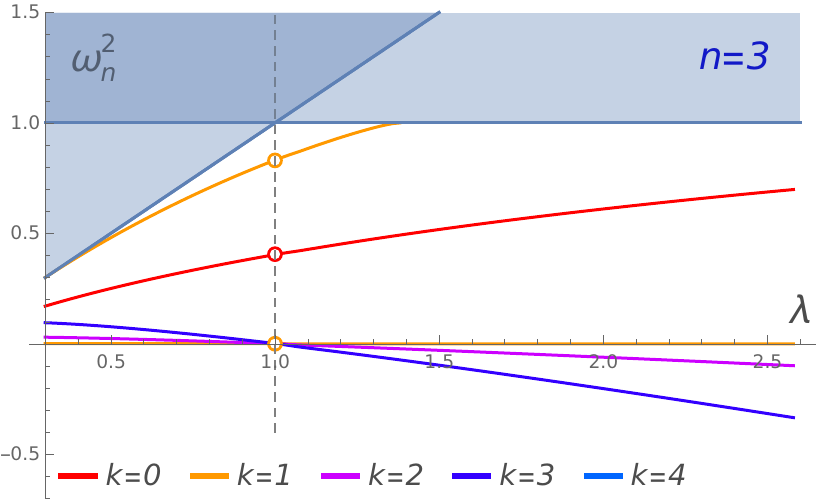} 
	\caption{Spectral structure for the rotationally invariant $3$-vortex as a function of the coupling constant $\lambda$. }  
	\label{fig:espectra3}
\end{figure}

\begin{figure}
	\centering
 \includegraphics[height=5cm]{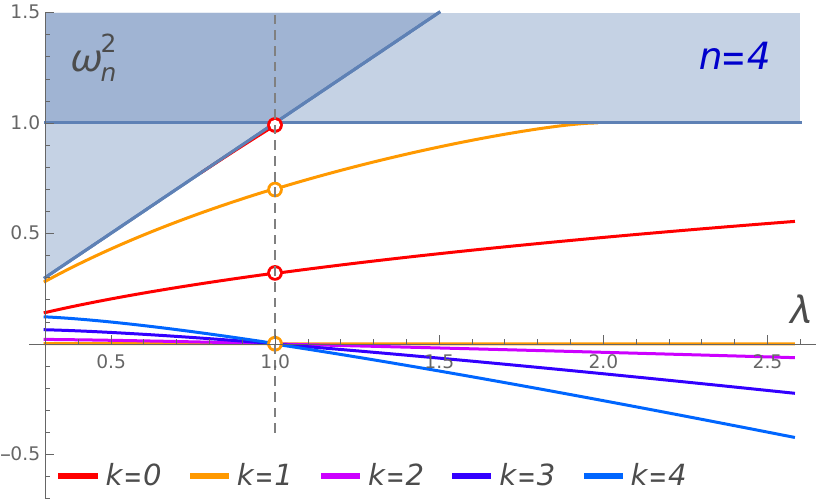} 
	\caption{Spectral structure for the rotationally invariant $4$-vortex as a function of the coupling constant $\lambda$. }  
	\label{fig:espectra4}
\end{figure}

In summary, the spectral problem (\ref{spectralproblem1}) determines both the zero modes and shape modes, except for the Derrick-type modes, which are governed by the spectral problem (\ref{spectralproblem0}). To explore the normal mode frequencies of an $n$-vortex for specific values of the coupling constant $\lambda$, a similar numerical scheme to that described in \cite{AlGarGuil,AlGarGuil2} can be employed on (\ref{spectralproblem0}) and (\ref{spectralproblem1}). Figures \ref{fig:espectra1}-\ref{fig:espectra4} illustrate the resulting spectral structure for the range $\lambda \in [0.3, 2.6]$. In our numerical analysis, the coupling constant $\lambda$ is varied in steps of $\Delta \lambda = 0.02$. In these figures, the known spectrum for the special case $\lambda = 1$ is indicated with hollow points, showing clear agreement between these points and the results of our general approach.

The continuum spectrum involves two distinct threshold values, corresponding to the scalar and vector fields. For the scalar field, the continuum spectrum begins at the value $\lambda$, while for the vector field, it starts at 1, as shown in Figures \ref{fig:espectra1}-\ref{fig:espectra4}.

In the case of 1-vortices, Figure \ref{fig:espectra1} shows the presence of one zero mode associated with $k=1$. Since this eigenvalue is doubly degenerate, the figure indicates the existence of two zero modes. Additionally, a Derrick-type mode (with $k=0$) is identified, which merges into the continuum spectrum around $\lambda=1.5$. For $k \geq 2$, the centrifugal barrier in the potential well of (\ref{spectralproblem1}) prevents the appearance of new bound states, and only the continuum spectrum is observed in these cases.

For self-dual 2-vortices, there are four zero modes. For general values of $\lambda$, only two of these zero modes (with $k=1$) persist, while the other two (with $k=2$), degenerate at $\lambda=1$, shift away from zero. These eigenvalues decrease as a function of $\lambda$, indicating that the 2-vortex is stable for $\lambda < 1$ but becomes unstable for $\lambda > 1$. A Derrick-type mode is also present and increases with $\lambda$. This behavior is general: eigenvalues related to eigenfunctions of type 2.1, which correspond to zero modes in the self-dual case, decrease with $\lambda$, while those related to type 2.2 (shape modes) increase. In Figure \ref{fig:espectra2}, a shape mode with $k=1$ is also visible near the threshold value at $\lambda \approx 1$.

The potential wells associated with the spectral problems become deeper as the vorticity $n$ increases, leading to a greater number of bound states in the spectrum. For $n=3$, the shape mode corresponding to $k=1$ becomes more prominent, and the degenerate zero modes at $\lambda=1$ split into three distinct curves, see Figure \ref{fig:espectra3}. A similar pattern is observed for 4-vortices, as shown in Figure \ref{fig:espectra4}.

\section{\label{sec:conclusions} Conclusions}

In this work, we have determined the spectrum of the second-order small fluctuation operator associated with rotationally invariant $n$-vortices in the Abelian Higgs model. By identifying the precise angular dependence of the eigenfunctions, we reduce the dimensionality of the original problem, making it more tractable and allowing for the eigenvalues to be computed with high precision. This knowledge of the eigenfunctions opens the door to studying the dynamics of excited vortices beyond the critical value $\lambda=1$. Such dynamics may alter the balance between attractive and repulsive forces in Type I and Type II superconductors, potentially leading to the surprising result that excited $n$-vortices could become stabilized in Type II superconductors.

\section*{Acknowledgements}

This research was funded by the Spanish Ministerio de Ciencia e Innovación (MCIN) with funding from the European Union NextGenerationEU
(PRTRC17.I1) and the Consejería de Educación, Junta de Castilla y León, through QCAYLE project, as well as the grant
and PID2023-148409NB-I00 MTM. D.M.C. acknowledges financial support from the European Social Fund, the Operational Programme of
Junta de Castilla y Leon and the regional Ministry of
Education.

\end{document}